\documentstyle[12pt,epsf]{article}

\fussy
\newcommand{\be}[1]{\begin{equation} \label{(#1)}}
\newcommand{\ee}{\end{equation}}
\newcommand{\ba}[1]{\begin{eqnarray} \label{(#1)}}
\newcommand{\ea}{\end{eqnarray}}

\newcommand{\beq}{\begin{equation}}
\newcommand{\eeq}{\end{equation}}
\newcommand{\beqa}{\begin{eqnarray}}
\newcommand{\eeqa}{\end{eqnarray}}

\newcommand{\kf}{k_{\mathrm F} }
\newcommand{\Sigs}{\Sigma_{\mathrm s} }
\newcommand{\Sigv}{\Sigma_{\mathrm v} }
\newcommand{\Sigo}{\Sigma_{\mathrm o} }

\newcommand{\bfgamma}{\mbox{\boldmath$\gamma$\unboldmath}}
\newcommand{\pabs}{|{\bf p}|}

\begin{document}
\title{Dirac Structure of the Nucleus-Nucleus Potential in Heavy Ion Collisions}
\author{T. Gross-Boelting, C. Fuchs and Amand Faessler\\
Institut f\"ur Theoretische Physik, 
Universit\"at T\"ubingen, D-72076 T\"ubingen, Germany}
\date{}
\maketitle
\begin{abstract}
We investigate nuclear matter properties in the relativistic Brueckner
approach. The in-medium on-shell T-matrix
is represented covariantly by five Lorentz invariant amplitudes from which we 
deduce directly the nucleon self-energy. 
To enforce correct Hartree-Fock results we develop a subtraction scheme 
which treats the bare nucleon-nucleon potential exactly in accordance to the 
different types of meson exchanges. 
For the higher order correlations
we employ two different covariant 
representations in order to study the uncertainty inherent in the approach. 
The nuclear matter bulk properties are only 
slightly sensitive on the explicit representation used.
However, we obtain new Coester lines for the various Bonn potentials 
which are shifted towards the empirical region of saturation.
\end{abstract}
\section{Introduction}
The investigation of nuclear matter properties within the
relativistic Dirac-Brueckner-Hartree-Fock (DBHF) approach 
\cite{horowitz87,terhaar87,brockmann90,sehn97,fuchs98,dejong98} 
remains a fundamental topic in theoretical nuclear structure studies. 
Compared to non-relativistic approaches the relativistic DBHF treatment 
turned out to be a major step forward in the explanation of the saturation
mechanism of nuclear matter. The saturation points obtained for 
non-relativistic calculations, throughout all possible choices of different
nucleon-nucleon interactions, are located on the so called 'Coester line'
\cite{coester70} which does not meet the empirical saturation region. 
Using modern nucleon-nucleon interactions of the one-boson exchange type 
\cite{erkelenz74} 
the relativistic calculations also reveal such Coester lines which are, however, 
significantly shifted towards the empirical region \cite{brockmann90}. 

On the other hand, many 
details of the relativistic theory are still not fully resolved. 
In particular, the precise form of the nucleon self-energy, i.e. 
the magnitude and the momentum dependence of the scalar and vector 
self-energy components are a question of current debate 
\cite{brockmann90,sehn97,fuchs98}. 
Since the self-energy describes the dressing 
of the particles inside the medium and thus determines the relativistic 
mean field this fact states a severe problem. Different 
techniques to handle the DBHF problem can lead to significantly different 
results \cite{brockmann90,sehn97,fuchs98,dejong98}. 
In a recent work \cite{fuchs98} we found that 
the momentum dependence of the nucleon 
self-energy is dominated by the one-pion exchange contribution which 
accounts for the nuclear tensor force. 

Unfortunately, the treatment of the $\pi$NN vertex and the corresponding
self-energy contributions is closely 
connected to a severe ambiguity in the T-matrix 
representation \cite{fuchs98}. 
The DBHF approach starts from a realistic nucleon-nucleon potential of the 
one-boson exchange type, i.e. the Bonn potentials \cite{machleidt86}.
As for the free two-body scattering problem, anti-particle states 
are neglected, and thus one works exclusively with positive energy states. 
Hence, a direct determination of the nucleon self-energy operator is 
not possible since not all matrix elements of this operator are known.
Horowitz and Serot have therefore developed a projection technique to 
determine the scalar and vector self-energy components 
from the in-medium T-matrix \cite{horowitz87}. 
In this approach the T-matrix is represented covariantly by 
Dirac operators and Lorentz invariant amplitudes where the latter are 
determined from the positive-energy on-shell T-matrix elements. 

The whole problem arises from the fact mentioned above, namely that one does not
include negative energy states and therefore neglects the excitation of 
anti-nucleons. The inclusion of negative energy excitations with 4 states for 
each spinor yields $4^4=256$ types of two-body matrix elements concerning their
spinor structure. Symmetry arguments reduce this to 44 for on-shell particles.
\cite{tjon85b}. If one takes now only positive energy solutions into account
this reduces to $2^4=16$ two-body matrix elements. Considering in addition only
on-shell matrix elements the number of independent matrix elements can be further
reduced by symmetry arguments down to 5. Thus, all on-shell two-body matrix elements
can be expanded into five Lorentz invariants. But these five invariants are not 
unique since the Dirac matrices involve always also negative energy states and thus
a decomposition of the one-body nucleon-nucleon potential into a Lorentz scalar and
a Lorentz vector contribution depends on the choice of these five Lorentz invariants
mentioned above. The best choice would be to separate completely the negative energy
Dirac states. But since this is not possible, there is not a unique but only an
'optimal choice'. The topic of this paper is the form of this 'optimal choice' of the
five invariants.

The paper is organized as follows: 
In section 2 we briefly review the Dirac-Brueckner Hartree-Fock approach.
In section 3 we introduce the projection technique and discuss the 
different covariant representations used for the on-shell T-matrix. 
Nuclear matter bulk properties are then discussed in section 4. 
At the end we summarize and conclude our work.
\section{The relativistic Brueckner approach}
\subsection{The coupled set of equations}
In the relativistic Brueckner approach the nucleon 
inside the nuclear medium is viewed as a dressed particle in consequence
of its two-body interaction with the surrounding nucleons. 
The in-medium interaction of the nucleons is treated in the ladder
approximation of the relativistic Bethe-Salpeter equation
\beq
T= V + i \int  VQGGT
\quad ,
\label{BSeq}
\eeq
where $T$ denotes the T-matrix. $V$ is the bare nucleon-nucleon interaction. 
The intermediate off-shell nucleons in the 
scattering equation are described by a two-particle propagator $iGG$.
We replace this propagator by the Thompson propagator \cite{thompson70}.
The Pauli operator $Q$ in the Thompson equation accounts for the 
influence of the medium by the Pauli-principle and projects the 
intermediate scattering states out of the Fermi sea. 

The Green's function $G$ fulfills the Dyson equation
\beq
G=G_0+G_0\Sigma G 
\quad .
\label{Dysoneq}
\eeq 
$G_0$ denotes the free nucleon propagator while the influence of the 
surrounding nucleons is expressed by the nucleon self-energy $\Sigma$. 
In Brueckner theory this self-energy is determined by summing up the 
interaction with all the nucleons inside the Fermi sea 
\beq
\Sigma = -i \int\limits_{F} (Tr[G T] - GT )
\quad .
\label{HFselfeq1}
\eeq
The Dirac structure of the self-energy in isospin saturated nuclear matter 
follows from translational and rotational invariance, parity conservation and 
time reversal invariance \cite{serot86}. 
In the nuclear matter rest frame the self-energy has the simple form 
\beq
\Sigma(k,\kf)= \Sigs (k,\kf) -\gamma_0 \, \Sigo (k,\kf) + 
{\bfgamma}\cdot {\bf k} \,\Sigv (k,\kf) 
\quad ,
\label{self1}
\eeq
with $k_\mu$ being the nucleon four-momentum. 
By taking the traces in Dirac space as \cite{horowitz87,sehn97}
\beq
\Sigs = \frac{1}{4} tr \left[ \Sigma \right] \quad ,\quad 
\Sigo = \frac{-1}{4} tr \left[ \gamma_0 \, \Sigma \right]
\quad , \quad 
\Sigv =  \frac{-1}{4|{\bf k}|^2 } 
tr \left[{\bfgamma}\cdot {\bf k} \, \Sigma \right] 
\label{trace}
\eeq
one can calculate the different Lorentz components of the self-energy.
\subsection{The in-medium T-matrix}
We apply the relativistic Thompson equation \cite{thompson70} to solve the 
scattering problem of two nucleons in the nuclear medium. 
In the two-particle center of mass (c.m.) 
frame - the natural frame for studying the two-particle 
scattering process - this Thompson equation can be written as 
\cite{terhaar87,sehn97}
\beqa
T({\bf p},{\bf q},x)|_{c.m.} &=&  V({\bf p},{\bf q}) 
\label{Tmateq}\\
&+& 
\int {d^3{\bf k}\over {(2\pi)^3}}
{\rm V}({\bf p},{\bf k})
{{\tilde m}^{*2}_F\over{\tilde{E}^{*2}({\bf k})}}
{{Q({\bf k},x)}\over{2{\tilde{E}}^*({\bf q})-2{\tilde{E}}^*({\bf k})
+i\epsilon}}
T({\bf k},{\bf q},x) 
\nonumber
\quad ,
\label{thompsoneq}
\eeqa
where ${\bf q}=({\bf q}_1 - {\bf q}_2)/2$ is the relative three-momentum 
of the initial state while $\bf k$ and $\bf p$ are the relative 
three-momenta of the intermediate and final states, respectively. 
The total four-momentum of the two-nucleon
system is $\tilde{P}^*=\tilde{q}^*_1+\tilde{q}^*_2$.
$\sqrt{\tilde{s}^*}=2{\tilde{E}}^*({\bf q})
=2\sqrt{{\bf q}^2+\tilde{m}^{*2}_F}$ is the starting energy in
(\ref{Tmateq}). If ${\bf q}_1$ and ${\bf q}_2$ are 
nuclear matter rest frame momenta of the nucleons in the initial state, 
the boost-velocity $\bf u$ into the c.m. frame is given by 
\beq
{\bf u} = {\bf P}/ \sqrt{\tilde{s}^{*}+{\bf P}^2}
\label{boost}
\quad ,
\eeq
with the total three-momentum and the invariant mass 
${\bf P} = {\bf q}_1+{\bf q}_2$ and $\tilde{s}^*=(\tilde{E}^*({\bf q}_1)
+\tilde{E}^*({\bf q}_2))^2-{\bf P}^2$, respectively.
In Eq. (\ref{thompsoneq}) x denotes the set of 
additional parameters $x=\{\kf, {\tilde m}^*_F,|{\bf u}|\}$ 
on which the T-matrix depends.

Applying standard techniques as explained in detail by Erkelenz \cite{erkelenz74}
we solve the Thompson equation in the c.m. frame and calculate the 
plane-wave helicity matrix elements of the T-matrix. 
The subspace of negative energy states is omitted in the current Brueckner 
approach. In this way we avoid the delicate problem of infinities in the
theory which would generally appear if we would include contributions 
from negative energy nucleons in the Dirac sea 
\cite{horowitz87,dejong98}. 
\section{Covariant representations and the nucleon self-energy}
\subsection{Pseudo-scalar representation}
\begin{figure}[h]
\vspace{75mm}
\includegraphics{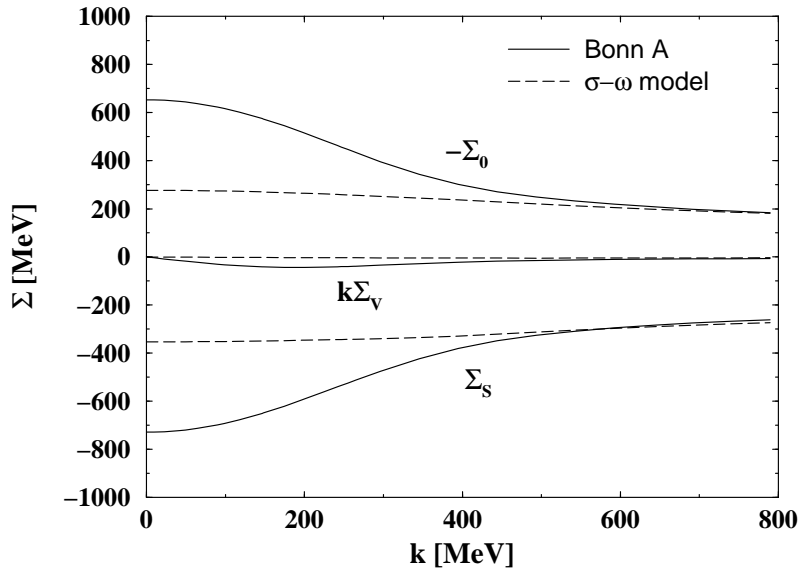}
\caption{\label{fig1} Momentum dependence of the DBHF nucleon self-energy
components in nuclear matter at $\kf=1.34 fm^{-1}$ using as bare 
nucleon-nucleon potential Bonn A (solid) and the $\sigma$-$\omega$ model 
potential (dashed). For the T-matrix the $ps$ representation 
(\ref{tmatrep1}) is applied.}
\end{figure}

To use the trace formulas, Eqs. (\ref{trace}), one has to represent the 
T-matrix covariantly. A set of five linearly independent covariants is 
sufficient for such a T-matrix representation because on-shell only 
five helicity matrix elements appear as solution of the Thompson equation. 
A linearly independent although not unique set of five covariants 
is given by the Fermi covariants 
\beqa
\!\!\!\!\!{\rm S} = 1\otimes 1 ,
{\rm V} =  \gamma^{\mu}\otimes \gamma_{\mu},
{\rm T} = \sigma^{\mu\nu}\otimes\sigma_{\mu\nu}, 
{\rm A} =  \gamma_5 \gamma^{\mu}\otimes \gamma_5 \gamma_{\mu},
{\rm P} = \gamma_5 \otimes \gamma_5 .
\eeqa
Using this special set - the so called 'pseudo-scalar choice' - 
the on-shell T-matrix for definite isospin I can be represented 
covariantly as \cite{horowitz87}
\beqa
\hspace{1cm}
T^{\rm I}(\pabs,\theta,x)&=& 
 F_{\rm S}^{\rm I}(\pabs,\theta,x){\rm S}
+F_{\rm V}^{\rm I}(\pabs,\theta,x){\rm V}
+F_{\rm T}^{\rm I}(\pabs,\theta,x){\rm T}
\nonumber \\
&+&F_{\rm A}^{\rm I}(\pabs,\theta,x){\rm A}
+F_{\rm P}^{\rm I}(\pabs,\theta,x){\rm P}
\label{tmatrep1}
\quad .
\eeqa
Here ${\bf p}$ and $\theta$ denote the relative three-momentum and the
scattering angle between the scattered nucleons in the c.m. frame, 
respectively. 
Applying the covariant representation (\ref{tmatrep1}) for the on-shell
T-matrix the nucleon self-energy in isospin saturated nuclear matter 
is evaluated to be \cite{sehn97}
\beq
\Sigma_{\alpha\beta}(k,\kf)= \int {{d^3{\bf q}}\over {(2\pi)^3}}
{{\theta(\kf-|{\bf q}|)}\over {\tilde{E}^*({\bf q})}}
\left[\tilde{m}^*_F 1_{\alpha\beta}F_{\rm S}
+\not{\tilde q}^*_{\alpha\beta}F_{\rm V}\right]
\quad ,
\label{self1}
\eeq
where the isospin averaged amplitudes are defined as
\beq
F_i(|{\bf p}|,0,x):=
{1\over 2}\left[ F_i^{{\rm I}=0}(|{\bf p}|,0,x)
+3 F_i^{{\rm I}=1}(|{\bf p}|,0,x)\right]
\quad .
\eeq
In Fig. \ref{fig1} we show the result of a self-consistent DBHF calculation for the 
nucleon self-energy in nuclear matter applying as representation for the
on-shell T-matrix the $ps$ representation (\ref{tmatrep1}).
As bare interaction we have used the Bonn A potential \cite{machleidt86} 
and, for comparison, the $\sigma$-$\omega$ model potential 
which was originally used by 
Horowitz and Serot \cite{horowitz87}. 
The Fermi momentum is $\kf=1.34 fm^{-1}$.
As already discussed in Ref. \cite{sehn97}, we see a pronounced 
momentum dependence of the nucleon self-energy 
components with the full Bonn A while in the case of the $\sigma$-$\omega$ 
model potential the dependence on the momentum is rather weak. 
A strong momentum dependence leads to unphysical
results deep inside the Fermi sea since the effective mass drops to
values which are close to zero. Therefore in Ref. \cite{fuchs98}
the strong momentum dependence of the self-energy was studied in more detail 
and found to originate mainly from the one-pion exchange contribution to the
self-energy. 

\subsection{Complete pseudo-vector representation}
\begin{figure}[h]
\vspace{75mm}
\includegraphics{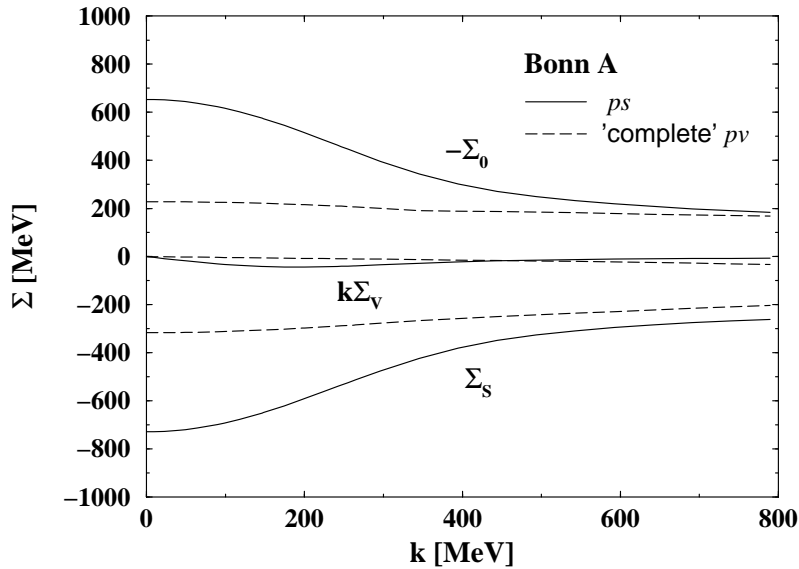}
\caption{\label{fig6} Momentum dependence of the DBHF nucleon self-energy
components in nuclear matter at $\kf=1.34 fm^{-1}$ 
using Bonn A as bare nucleon-nucleon interaction. For the 
T-matrix the $ps$ representation (Eq. (\ref{tmatrep1}), solid) and the 
'complete' $pv$ representation (Eq. (\ref{tmatrep6}), dashed) are applied.}
\end{figure}

To suppress the undesirable pseudo-scalar contribution of the pion to the 
nucleon self-energy we have to use the 'complete' $pv$ representation 
of the T-matrix \cite{tjon85a}
\beqa
\hspace{1cm}
T^{\rm I}(\pabs,\theta,x)&=& g_{\rm S}^{\rm I}(\pabs,\theta,x){\rm S}
-g_{\tilde{\rm S}}^{\rm I}(\pabs,\theta,x)\tilde{\rm S}
+g_{\rm A}^{\rm I}(\pabs,\theta,x)({\rm A}-\tilde{\rm A})\nonumber \\
&+& g_{\rm PV}^{\rm I}(\pabs,\theta,x){\rm PV}
-g_{\widetilde{\rm PV}}^{\rm I}(\pabs,\theta,x)\widetilde{\rm PV}
\quad .
\label{tmatrep6}
\eeqa
The amplitudes $g^{\rm I}(\theta)$ are explicitly given in \cite{gross98}.
In Fig. \ref{fig6} we present the full self-consistent DBHF calculation 
with the 'compete' $pv$ representation of the T-matrix \cite{fuchs98}. 
The DBHF nucleon self-energy components are indeed weakly momentum dependent.
The single-pion exchange contribution to the interaction, 
which was previously dominating at low nucleon momenta, 
is now strongly suppressed.  
Consequently the result within the 'complete' pv representation using the
Bonn A potential resembles the result within the $\sigma$-$\omega$ model 
potential, see Fig. \ref{fig1} where the $ps$ representation was used. 
To suppress the pion contribution to the in-medium T-matrix 
a correct pseudo-vector like covariant representation is essential for the 
calculation of the nucleon self-energy in nuclear matter. 
\subsection{Covariant representations of the subtracted T-matrix}
The 'complete' $pv$ representation successfully 
reproduces the HF nucleon self-energy in the case of the pion exchange.
However, as already pointed out in \cite{fuchs98}, the 'complete' 
$pv$ representation fails to reproduce the HF nucleon self-energy if other 
meson exchange potentials are applied as bare interaction. 
Hence, it should be reasonable to treat the bare interaction and 
the higher order ladder graphs of the meson exchange potential separately. 
Since the single-meson exchange potential is actually known analytically 
we can represent it covariantly by a mixed representation of the form 
\beq
V=V_{\pi,\eta}^{PV}+V_{\sigma,\omega,\rho,\delta}^{P}
\quad .
\label{mixed1}
\eeq
Here the $\pi$- and $\eta$-meson contributions  
are treated as pseudo-vector 
while for the ($\sigma,\omega,\rho,\delta$)-meson contributions 
of the Bonn potential the $ps$ representation is applied. 
The higher order ladder diagrams of the T-matrix 
\beq
T_{Sub}=T-V=i\int VQGGT = \sum\limits_{n=1}^\infty \int V(iQGGV)^n
\quad ,
\eeq
in the following called the subtracted T-matrix, can not be represented 
correctly in a mixed form since we can not disentangle the different
meson contributions to this part of the full in-medium interaction.
The representation of the subtracted T-matrix remains therefore ambiguous.
However, if the pion exchange dominantly contributes to 
the Hartree-Fock level a $ps$ representation of the subtracted T-matrix 
should be more appropriate because then the higher order contributions of 
other meson exchange potentials are not treated incorrectly as pseudo-vector. 
Thus the most favorable representation of the T-matrix is given by the 
$ps$ representation
\beq
T^{\rm P}=T^{\rm P}_{Sub}+V_{\pi,\eta}^{PV}+V_{\sigma,\omega,\rho,\delta}^{P}
\quad .
\label{tmatps}
\eeq
Here the $ps$ representation for $T^{\rm P}_{Sub}$ is determined via the 
matrix elements
\beqa
<{\bf p}\lambda_1^{'}\lambda_2^{'}|T^{\rm I}_{Sub}(x)|
{\bf q}\lambda_1\lambda_2>:=
<{\bf p}\lambda_1^{'}\lambda_2^{'}|T^{\rm I}(x)-V^{\rm I}(x)|
{\bf q}\lambda_1\lambda_2> 
\quad ,
\label{tmatsub}
\eeqa
with subsequently applying the projection scheme as in Eq. (\ref{tmatinv}).
An alternative representation of the T-matrix is given
by the $pv$ representation
\beq
T^{\rm PV}=T^{\rm PV}_{Sub}+V_{\pi,\eta}^{PV}+V_{\sigma,\omega,\rho,\delta}^{P}
\quad ,
\label{tmatpv}
\eeq
where the subtracted T-matrix is represented by the 'complete' 
$pv$ representation (\ref{tmatrep6}). This representation is similar
to the 'complete' $pv$ representation of the full T-matrix, however, with the
advantage that now the pseudo-scalar contributions in the bare 
nucleon-nucleon interaction, e.g. the single-omega exchange potential, 
are represented correctly. In the next section we will use both 
representations, (\ref{tmatps}) and (\ref{tmatpv}), to study the properties 
of nuclear matter in the DBHF approach. 
In this way we can determine the influence of the higher order 
ladder graphs to the in-medium interaction in a more quantitative way. 
Furthermore, these two representations set the range of the remaining 
ambiguity concerning the representation of the T-matrix, i.e. after
separating the leading order contributions.
\section{The equation-of-state of nuclear matter}
\begin{figure}[h]
\vspace{75mm}
\includegraphics{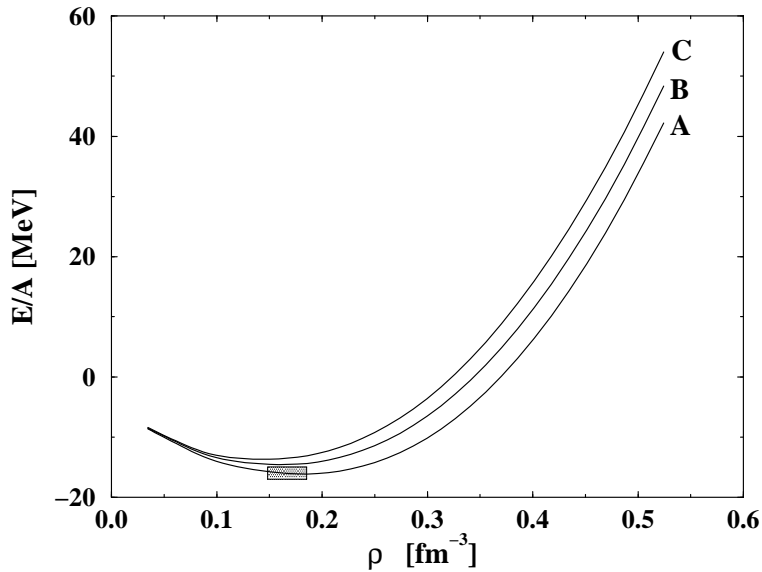}
\caption{\label{fig13} Binding energy per particle as a 
function of nuclear matter density. 
As bare nucleon-nucleon interaction the potentials 
Bonn A, B and C are used. For the T-matrix the subtraction scheme with the 
$ps$ representation (\ref{tmatps}) is applied. 
The shaded box denotes the empirical saturation region of nuclear matter.}
\end{figure}

In the relativistic Brueckner theory the energy per particle is defined 
as the kinetic plus half the potential energy 
\beq
E/A  =  {1\over \rho}\sum_{{\bf k},\lambda}
<\bar{u}_\lambda({\bf k})| 
\bfgamma \cdot {\bf k} + M + {1\over 2}\Sigma (k)
| u_\lambda({\bf k})>
{{\tilde{m}^*(k)}\over {{\tilde{E}}^*(k)}} - M
\quad .
\label{eos}
\eeq
In Fig. \ref{fig13} we show the binding energy per particle $E/A$ as a 
function of the density, calculated with Bonn A, B and C.
For the T-matrix the subtraction scheme with the $ps$ representation 
(\ref{tmatps}) is applied.
With Bonn A one can reproduce the empirical saturation point
of nuclear matter, shown as shaded region in the figure.
The other Bonn potentials give less binding energy although the 
saturation density is always close to the empirically known value.
The result for the binding energy per particle using the two 
representations (\ref{tmatps}) and (\ref{tmatpv}) for the T-matrix 
are very similar.
At saturation density the binding energy is only 0.5 MeV smaller using the 
pseudo-vector representation of the subtracted T-matrix. 
Thus, the energy per particle is not very sensitive on the explicit representation of the 
subtracted T-matrix. 
\begin{figure}[h]
\vspace{75mm}
\includegraphics{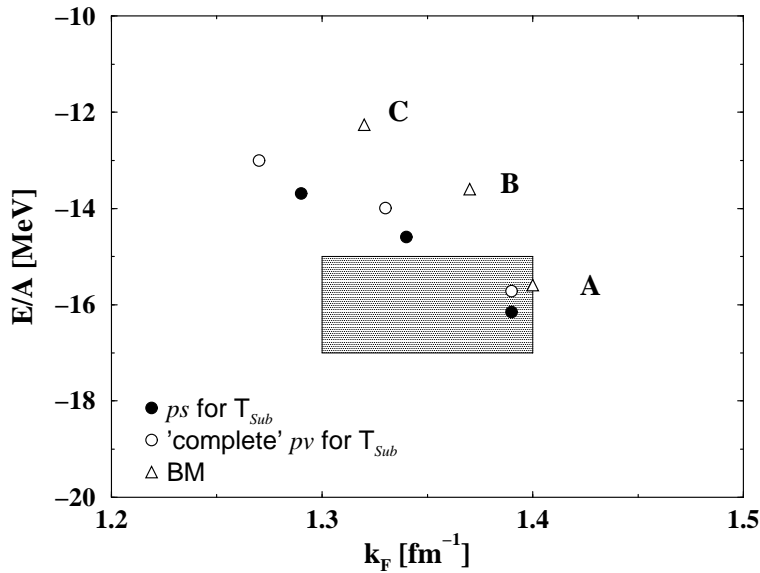}
\caption{\label{fig16} Saturation points of nuclear matter. As bare 
nucleon-nucleon interaction the Bonn potentials A,B and C are used.
For the T-matrix the subtraction scheme with the 
$ps$ representation (Eq. (\ref{tmatps}), filled circles) 
and the $pv$ representation (Eq. (\ref{tmatpv}), open circles) 
are applied. As open triangles the results of the calculation of 
Brockmann and Machleidt (BM), Ref. \protect\cite{brockmann90}, are shown.}
\end{figure}

The present results are summarized in Fig. \ref{fig16} 
where the corresponding saturation points for the three 
different versions of the Bonn potential are shown. 
We compare the results with the two representation of the
subtracted T-matrix with the results of the calculation of 
Brockmann and Machleidt (BM), Ref. \cite{brockmann90}.
With the improved representation schemes 
(\ref{tmatps}) and (\ref{tmatpv}) for the T-matrix one obtains new 
'Coester-lines' which are left of the original one, 
i.e. shifted towards the empirical region. The refined treatment of the 
T-matrix representation leads to an enhancement
of the binding energy connected with a reduced saturation density. As
in the previous calculations, Bonn A is still the only one which 
meets the empirical region. However, due to an increased binding 
energy Bonn B is now much closer to empirical region. This observation 
is consistent with the present treatment. The different types of Bonn 
potentials essentially vary in the strength of the nuclear tensor 
force determined by the $\pi NN$ form factor. Bonn A which has the 
smallest tensor force yields the smallest D-state probability of 
the deuteron and only a pure description of the 
$^{3}D_{1}$ phase shift \cite{brockmann90,machleidt86}. 
Thus it appears that a refined treatment of the 
pion exchange leads to improved nuclear matter results for the 
more realistic Bonn B potential. Bonn C, however, is still 
far off the empirical region.

Furthermore, it can be seen from Fig. (\ref{fig16}) 
that the final nuclear matter bulk properties depend 
only moderate on the representation of the 
subtracted T-matrix. In Ref. \cite{fuchs98} we tried already to 
determine the range of inherent uncertainty in the relativistic 
Brueckner approach which is due to the ambiguities concerning the 
representation of the T-matrix discussed in Section 3. 
By the separate treatment of the Born contribution to the T-matrix 
we end up now with a much narrower uncertainty band which is 
given by the $ps$ or complete $pv$ representation of the ladder 
kernel, i.e. the subtracted T-matrix. Over the different types 
of Bonn interactions the two methods lead to a variation of 
0.5 MeV in the binding energy, 0.1--0.2 $fm^{-1}$ in the Fermi 
momentum, and to about 30 MeV concerning the value of the 
effective mass at saturation density. The values for 
incompressibility are also close in the two approaches, i.e. they 
differ by less than 10 MeV.  
Within the $ps$ representation of the subtracted T-matrix, 
Bonn B and C now yield very small kompression moduli around
$K=150 MeV $ and $K=115 MeV$, respectively. 
For Bonn A a value of $K=230 MeV$ is obtained. 
This value agrees with the empirical value of the 
kompression modulus of $K=210 \pm 30 MeV$.
Here Brockmann and Machleidt found much larger values for all three 
Bonn potentials. 
\section{Summary}
We have investigated the nuclear matter properties in the 
relativistic Brueckner approach. The required representation of 
the T-matrix by Lorentz invariant amplitudes suffers thereby 
from on-shell ambiguities concerning the pseudo-scalar or 
pseudo-vector nature of the interaction. We minimized this ambiguity 
by separating the leading order, i.e. the single-meson exchange,  
from the full T-matrix. 
The remaining higher order correlations, i.e. the ladder kernel, are then 
represented either completely as pseudo-scalar or as pseudo-vector. 

As a major result of our investigation we obtain new 'Coester lines' 
for the various Bonn potentials. Compared to previous treatments 
these are shifted towards
the empirically know saturation point. Bonn A is still the only potential
which really meets the empirical region of saturation, but, 
with improved saturation properties compared to 
previous treatments. The refined treatment of the pion exchange 
leads on the other hand also to improved results for 
the -- from the view of the phase shift analysis -- more realistic 
Bonn B potential. Furthermore, we found that the 
equation-of-state is strongly softened compared to previous calculations.
Actually with Bonn A we obtain a kompression modulus of $K\sim 230 MeV$ 
which is in good agreement with the empirical value.

To summarize our results, we obtained new results for the nuclear matter
properties within the projection technique employing an new method for the 
T-matrix representation. The final results are at lower
densities almost insensitive on the explicit choice made for the 
representation. However, at higher densities, certain differences occur when 
using different representation schemes. We want to stress that the ambiguity in the
projection technique is still not fully resolved yet. We plan to look on off-shell
T-matrix elements in the future since off-shell matrix elements of the 
pseudo-scalar and pseudo-vector covariants differ significantly. 
We hope that this might bring more insight on what is the correct on-shell 
representation of the T-matrix. 

\end{document}